\documentclass[11pt]{article}%
\usepackage{picinpar,moriond,epsfig}
\def\Journalbis#1#2#3#4{{#1} {\bf #2} (#3) {#4}}%
\def\AP{\em Astroparticle Physics}
\def\NIMA{{\em Nucl. Instrum. Methods} A}%
\def\NPBP{{\em Nucl. Phys.} B (Proc. Suppl.)}
\def\PRL{\em Phys. Rev. Lett.}%
\def\be{\begin{equation}}%
\def\ee{\end{equation}}%
\def\bea{\begin{eqnarray}}%
\def\eea{\end{eqnarray}}%
\begin{document}%
\vspace*{4cm}%
\title{STATUS OF THE BOREXINO EXPERIMENT}%
\author{T.J. BEAU}%
\address{Physique Corpusculaire et Cosmologie\\
Coll{\`e}ge de France, Paris, France}%
\maketitle
\abstracts{
A short review of the solar neutrino Borexino experiment is given.}
%
Borexino is a solar neutrino experiment which is in its final step of completion in the underground laboratory at Gran Sasso, Italy. 
It will detect in real time the low energy part of the  solar neutrino spectrum (below 1~MeV); 
thus, it will give new contraints to the solar neutrino problem, after the important SNO result\cite{sno} in june 2001.
%
\section{Borexino Physics goal}
The main goal of Borexino is the measurement of the $\nu_{^7\textrm{Be}}$ flux from the Sun in real time. In few time, we will be able to know if a LOW solution is possible by the day-night asymetry. 
Moreover, Borexino will be sensitive to any seasonal variation effect and will give a precise flux for the monoenergetic ray of the $\nu_{^7\textrm{Be}}$ (862~keV). The expected rate for this flux is around 40~counts/day, from the Standard Solar Model (SSM).

In addition, Borexino will also be concerned in other solar neutrinos (Boron, CNO), nuclear reactor neutrinos\cite{react} or supernovae neutrinos\cite{sn}.

%
\section{The detector}
\subsection{Detection principle and detector description}
Borexino\cite{bxsc} detects all flavor neutrinos by electron elastic diffusion, $\nu_x e\to\nu_x e$. If $\nu_e$ is driven by charged current (CC)  and by neutral current (NC) scattering, $\nu_{\mu}$ and $\nu_{\tau}$ are only driven by NC. The event rate relative to NC will be only about 20 percent of the total rate.

The detector has a shell structure (see figure \ref{fig:bx}).
From outside to the center of the detector, one can find water shield, first pseudocumen (PC) buffer (not scintillating), nylon film as a Rn barrier, second pseudocumen buffer, nylon film and scintillating pseudocumen (300 tons).
100 tons are defined by offline analysis as the fiducial volume. 
Scintillation photons are detected by 2200 photomultiplier tubes (PMTs) which will work as single photoelectron (pe) detectors for low energy events ($<1.5\:\rm{MeV}$). The expected photoelectron yield is about 450~pe/MeV.

The Borexino detector is located at the National Laboratory of Gran Sasso (LNGS), Italy. 
This underground laboratory provides a 3800 meters water equivalent shield; only one muon per hour per square meter arrives at this depth.

\subsection{Schedule}
\begin{figwindow}[0,r,%
{\fbox{\epsfig{file=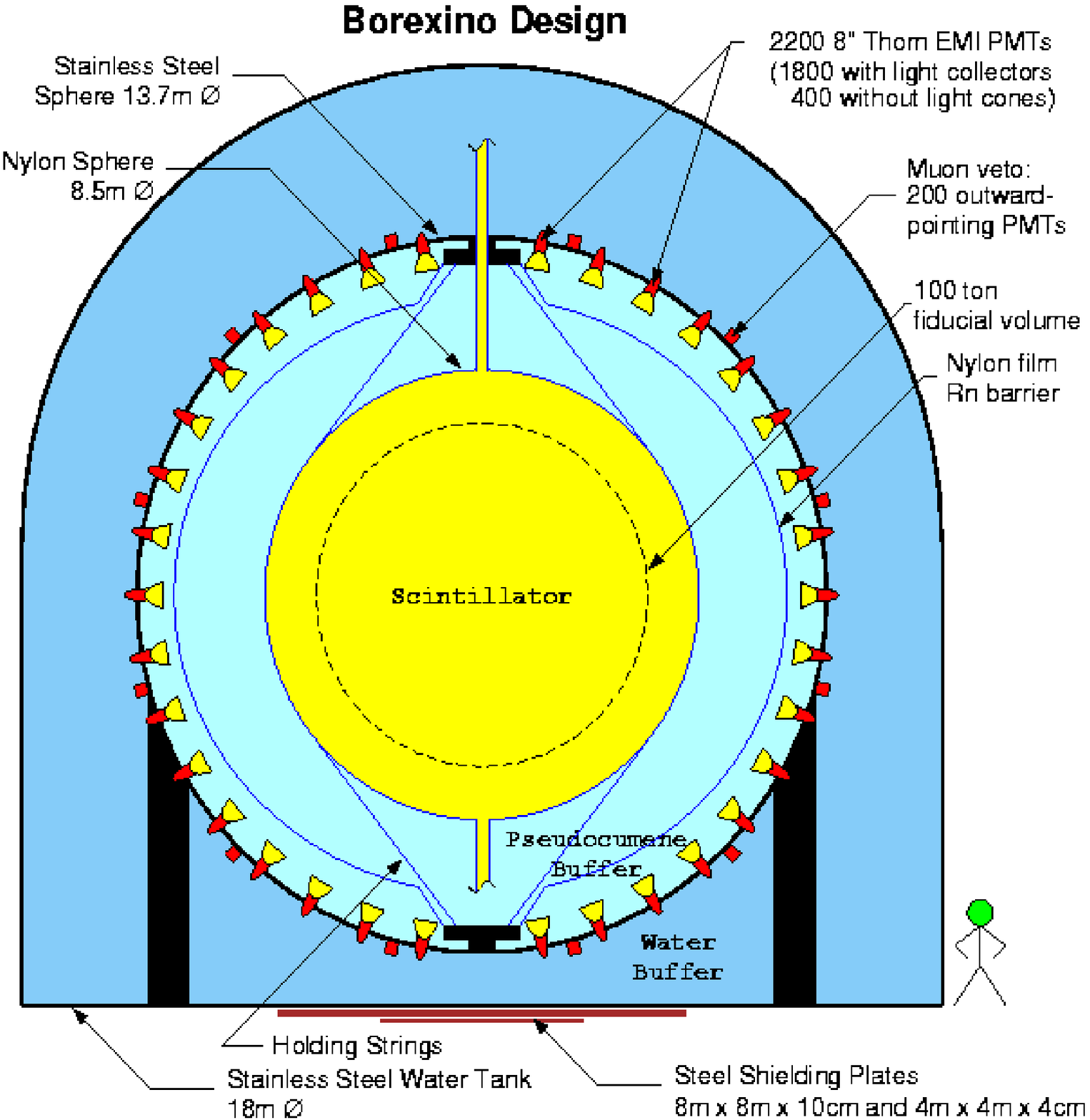,width=70mm}}},%
{Borexino detector\label{fig:bx}}]
During previous years, a very big effort was made to achieve an extremely low radioactivity level for Borexino in order to be able to perform low energy neutrino spectroscopy\cite{lowbg}. 
The Counting Test Facility\cite{ctf} (CTF), a multi-ton scintillator detector, was designed in order to analyse radiopurity of such a detector and to tune all parameters needed for Borexino.
There have been three main steps in CTF, where different nylon vessels and scintillators have been tested.

Here are the main steps of the foreseen Borexino schedule~: \\
$\bullet$ June 2002~: Nylon vessel installation, \\
$\bullet$ Septembre 2002~: Water Filling, \\
$\bullet$ Automn 2002~: PC Filling (and background measurement),  \\ 
$\bullet$ April 2003~: Borexino ready for neutrino physics.
\end{figwindow}
\vspace{4mm}
\section{Electronics and data acquisition (DAQ)}
The outer detector is a 200 PMT muon veto system. The inner detector DAQ can be basically divided into two parts~:

$\bullet$ An unique photoelectron electronics, optimized for the $^7\rm{Be}$ energy window (250--800~keV). There is one channel per PMT, digitalizing time and charge.

$\bullet$ A 400 MHz Flash ADC system, based on VME boards designed in our laboratory, which will soon be available as CAEN V896. 
Those boards have no dead time, 10 $\mu\rm{s}$ (programmable size) window per event, 64 event buffer storage. 
99 channels will be used, each of them digitalizing roughly 24 PMT signals. 
The design of the DAQ software allows a massive data reduction. 
This electronics gives the main information for non-centered events and physics above 1~MeV such as nuclear reactor neutrinos, $^8\rm{B}$ and CNO solar neutrinos, supernovae neutrinos.
\section*{References}%
\end{document}